\documentstyle[twocolumn,epsf]{jpsj}
\catcode`\@=11

\def\simle{\mathrel{\mathpalette\@versim<}}   % < over \sim
\def\simge{\mathrel{\mathpalette\@versim>}}   % > over \sim
\def\@versim#1#2{\lower2.5pt\vbox{\baselineskip0pt \lineskip-.5pt
   \ialign{$\m@th#1\hfil##\hfil$\crcr#2\crcr\sim\crcr}}}
\catcode`\@=12
\def\runtitle{Order $N$ Monte Carlo Algorithm
}
\def\runauthor{ Nobuo {\sc Furukawa} and Yukitoshi {\sc Motome}}

\def\sf{\rm}

\title{
Order $N$ Monte Carlo Algorithm for Fermion Systems\\
Coupled with Fluctuating Adiabatical Fields
}

\author{Nobuo {\sc Furukawa} and Yukitoshi {\sc Motome}$^1$
}
\inst{
Department of Physics, Aoyama Gakuin University, 
Fuchinobe 5-10-1, Sagamihara, Kanagawa 229-8558\\
$^{1}$RIKEN (The Institute of Physical and Chemical Research),
2-1 Hirosawa, Wako, Saitama 351-0198 
}
\recdate{\today}

\abst{
An improved algorithm is proposed
for Monte Carlo methods to study  fermion systems
interacting with adiabatical fields.
To obtain a weight for each Monte Carlo sample
with a fixed configuration of adiabatical fields,
a series  expansion using Chebyshev polynomials
is applied.
By introducing truncations of matrix operations in a systematic
and controlled way,
it is shown that the cpu time is reduced from
$O(N^3)$ to $O(N)$
where $N$ is the  system size.
Benchmark results show that the implementation of the 
algorithm makes it possible 
to perform  systematic investigations of critical phenomena
 using system-size scalings
even for an electronic model in
 three dimensions, within a realistic cpu timescale.
}

\kword{
Monte Carlo method, Order $N$ method, moment expansion, Chebyshev polynomials,
parallel computation, double-exchange model, critical phenomena
}

\begin{document}
\sloppy
\maketitle

%%%%%%%%%%%%%%%%%%%%%%%%%%%%%%%%%%%%%%%%%%
\section{Introduction}
There exists a lot of interests
in the class of strongly correlated electron systems
where itinerant electrons are coupled to adiabatical fields.
An example is the strongly coupled electron-lattice system, while the other
is the double-exchange system where electrons 
are coupled to localized classical spins.
Models which represent dilute magnetic semiconductors
also belong to this class.

Ground states of such systems may be
studied by mean-field approaches, since fluctuations
are frozen and irrelevant.
However, in order to study finite temperature
properties, especially near the critical temperature,
it is necessary to take into account fluctuations of the fields.
Natures of the field fluctuations
are important to
understand  critical phenomena as well as changes of
 electronic properties around critical points.

Since dynamics of the fluctuations are adiabatically slow,
electrons can respond to the fluctuating potentials
precisely so that the electronic states are far from
those without fluctuations.
Therefore, in the presence of critical fluctuations 
which are strongly coupled to
electrons, various theoretical methods such as mean-field approaches
and perturbation theories are invalid.

Numerical studies provide promising methods to
calculate such systems.
Especially,
Monte Carlo (MC) method
is suited for calculation of these models.
The advantage of this method is that it is possible to obtain
 thermodynamics of the model on a finite size lattice
by taking partition sums for fluctuating fields which are
replaced by stochastic samplings.\cite{Yunoki98}

However,  MC studies suffer from finite-size effects
since the system size is  limited, due to an increase
of the computational complexities and hence cpu time
as system sizes are increased.
In order to study thermodynamic properties of the model properly,
it is requisite to perform 
extrapolations to the thermodynamic limit
as well as finite size scalings.
Namely, systematic calculations for various lattice sizes
which are large enough for analyses are necessary.
In the conventional algorithm,\cite{Yunoki98}
the computational complexity scales as $O(N^4)$,
where $N$ is the system size.
Therefore, it is extremely difficult to increase the system size.

In order to overcome the difficulty, 
improved algorithms have been proposed
 to reduce the computational complexity
for the calculation of the Boltzmann weight.
The authors have introduced 
the polynomial expansion method (PEM), where
the computational complexity is reduced to $O(N^3)$.\cite{Motome99b,Motome00we}
Alonso {\em et al.}\cite{Alonso01} have applied 
the hybrid MC algorithm which makes the
computational complexity to scale as $O(N^2)$.
Using these new methods, larger system sizes become 
available.
As an example, critical phenomena of the two-dimensional
double-exchange model have been investigated using
finite-size scaling analysis and non-equilibrium
relaxation studies.\cite{Motome01,Motome01b}
However, these algorithms are still not sufficient enough
to study systems which require much larger computational scales,
such as models with more complex interactions or
those in three dimensions.

In this paper, we present a new algorithm which
further reduces the computational complexity
of the MC calculations to $O(N)$.
In \S2, we briefly describe the PEM
in order to make this article self-contained.
In \S3, we introduce a truncation method to
improve the PEM.
Benchmark results and estimates of the truncation errors
are shown in \S4. Sec. 5 is devoted to summary and discussions.

\section{Polynomial Expansion Monte Carlo Method}
\label{Section:PEM}

\subsection{Hamiltonian matrices}
Throughout this paper, we consider a system where
the Hamiltonian operator is expressed in a quadratic form,
\begin{equation}
 \hat {\cal H}({\mib\phi}) = \sum_{ij}  c_i^\dagger H_{ij}({\mib\phi}) c_j.
  \label{defHgeneral}
\end{equation}
Here, $c_i$ ($c_i^\dagger$) represents a fermion annihilation (creation)
operator for an index $i$. Each index represents  fermionic degrees of 
freedom, typically a combination of site, orbital and spin.
Matrix elements depend on the adiabatical fields
which are expressed by $\mib\phi$.
This means that we restrict ourselves to
 a class of electronic systems on lattices
coupled to adiabatical fields which give,
{\em e.g.}, charge/spin density potentials as well as
those coupled to orbital degrees of freedom as in Jahn-Teller couplings.
We assume  the absence of  electron-electron interactions. 

Later in this paper, simple examples of the Hamiltonian (\ref{defHgeneral})
will be given in Eq.~(\ref{defHmatrixExample}), 
where spinless electrons are
coupled to on-site potential fields, and in Eq.~(\ref{defHDE})
where electrons are coupled to localized spins
defined on each site in such a way that
transfer energies are modulated by configurations of the spins.

Let us describe the class of systems which are
expressed by the Hamiltonian (\ref{defHgeneral}) more precisely.
In usual cases,
diagonal elements of the Hamiltonian matrices describe 
potential energies, {\em e.g.}, charge density potentials
coupled to  adiabatical fields.
Electron hopping terms,
as well as coupling to
transverse fields give  off-diagonal matrix elements of $H$.
Within the scheme
 we do not consider types of fields which break electron number
conservation,
{\em e.g.}, coupling to singlet superconducting fields in a form
$ \Delta_{i} c_{i\uparrow}^\dagger c_{i\downarrow}^\dagger + h.c.$,
unless there exists a canonical transformation which
maps the system to a particle-number conserving system,
{\em e.g.}, $c_{i\uparrow} \to d_{i\uparrow}$ and $c_{i\downarrow} \to
d_{i\downarrow}^\dagger$ in the previous example.

For a fixed configuration of the adiabatical fields,
the Hamiltonian in Eq.~(\ref{defHgeneral}) shows 
a one-body electron system with
random potentials.
The Hamiltonian matrices $H$ have a matrix dimension $N_{\rm dim}$ 
defined by the total number of fermionic degrees of freedom,
which is proportional to
the system size $N$.
Within this article we restrict ourselves to the cases where
$H$ are sparse matrices, namely, the model has
short range hoppings in usual cases.

 We assume that the adiabatical fields are locally defined,
typically on sites or bonds. Then, the total number of the adiabatical fields
is proportional to $N$.
We also restrict ourselves to 
the case where interactions between the fields 
and the electrons are short-ranged,
{\em i.e.}, the number of matrix elements which are
modulated by the change of an adiabatical field is $O(N^0)$.

\subsection{Boltzmann weight}
The partition function for the Hamiltonian
(\ref{defHgeneral}) is written as
\begin{eqnarray}
\label{Zdef}
  Z &=& {\rm Tr_{C}}    {\rm Tr_{F}} \exp \left( -\beta \left[
	\hat{\cal H} \left( {\mib\phi} \right) - \mu \hat{N}_{\rm e}
	\right] \right),
\end{eqnarray}
where ${\rm Tr_{C}}$ is the trace over
adiabatical fields $\mib\phi$, while ${\rm Tr_{F}}$ is the
grand canonical trace over fermion degrees of freedom.
Here, $\beta$ is the inverse temperature, $\mu$ is the chemical 
potential and $\hat{N}_{\rm e}$ is the particle-number operator.

In MC approaches, the trace over adiabatical fields
is replaced by the stochastical sampling of the
field configurations ${\mib\phi}$
whose Boltzmann weight is given by
\begin{eqnarray}
  P({\mib\phi}) = \frac1{Z}
  \exp \left[ -S_{\rm eff}\left({\mib\phi}\right) \right].
  \label{defP}
\end{eqnarray}
Here, the effective action $ S_{\rm eff}$ is defined by
\begin{eqnarray}
%  \label{defSeff0}
   S_{\rm eff}\left({\mib\phi}\right)
 &\equiv& 
 -\log\left({\rm Tr_{F}} \,{\rm e}^{  -\beta \left[
	\hat{\cal H} \left( {\mib\phi} \right) - \mu \hat{N}_{\rm e}
	\right] } \right)
   \nonumber \\
 &=&
    \sum_{\nu=1}^{N_{\rm dim}} F\left( 
            \varepsilon_{\nu} ({\mib\phi} ) \right),
  \label{defSeff}
\end{eqnarray}
where
\begin{eqnarray}
  F(x) &=& - \log \left[ 1 + {\rm e}^{-\beta(x-\mu)}\right],
  \label{defFx}
\end{eqnarray}
while $\varepsilon_\nu$ is the $\nu$-th eigenvalue of the
Hamiltonian matrix $H$ for a given configuration of $\mib\phi$.

In an importance sampling MC method, probability of an update
from an old field configuration ${\mib\phi^{\rm old}}$
to a new configuration ${\mib\phi^{\rm new}}$ depends on
the ratio of the Boltzmann weights
which is given by
\begin{equation}
 r = \frac{P({\mib\phi^{\rm new}})}{P({\mib\phi^{\rm old}})}.
  \label{defBoltzRatio}
\end{equation}
In a local update of the adiabatical fields,
we generate $\mib \phi^{\rm new}$ from $\mib\phi^{\rm old}$
in such a way that only one or a local group of adiabatical fields
is modified from $\mib\phi^{\rm old}$ but the rest are unchanged.
The definition of a MC step with local updates
is that we make a sweep of local updates 
so that all adiabatical field variables are sequentially
 examined for updates.

One of the method to calculate the Boltzmann weight 
$P$ from Eqs.~(\ref{defP})-(\ref{defFx}) 
is the diagonalization method (DM)
where $ \varepsilon_{\nu} ({\mib\phi} )$ are exactly obtained by
direct diagonalizations of the 
Hamiltonian matrix $H({\mib\phi} )$.\cite{Yunoki98}
The computational complexity for each matrix diagonalization
to obtain all eigenvalues
is $O(N_{\rm dim}{}^3)$. 
In a MC step with local updates, the number of trials for the field upgrades
scales as $O(N)$. 
Since $N_{\rm dim} \propto N$, the total computational
 complexity for a MC step by the DM is $O(N^4)$. 

\subsection{Polynomial expansion method}

An approach to reduce the computational complexity in the calculation 
of $P({\mib\phi})$
is to perform a polynomial expansion.\cite{Motome99b}
When $F(x)$ in Eq.~(\ref{defFx}) is expanded by a series of polynomials
$\{ T_m(x)\}$ in a form 
\begin{equation}
  \label{defPolyExpand}
  F(x) = \sum_{m=0}^\infty f_m T_m(x),
\end{equation}
we may rewrite Eq.~(\ref{defSeff}) as
\begin{equation}
  \label{CalcSeff}
    S_{\rm eff}(\mib \phi)  = \sum_m f_m \,\mu_m,
\end{equation}
where $\mu_m$ is the polynomial moments of the
Hamiltonian defined by
\begin{equation}
  \mu_m  = \sum_{\nu=1}^{N_{\rm dim}} T_m(\varepsilon_\nu(\mib \phi)) 
 = {\rm Tr}\, T_m(H(\mib \phi)).
  \label{defmu}
\end{equation}
Here ${\rm Tr}$ represents
a trace operation for the matrix polynomials.
The values of  $f_m$ depend on temperature and chemical potential
but not on the adiabatical fields $\mib\phi$.
Moments  $\mu_m$ depend on $\mib\phi$, which
 are calculated through  trace operations
and matrix addition/multiplications of $H(\mib\phi)$.
Therefore, the  expansion of the effective
action $S_{\rm eff}$ by
a polynomial series enables us to obtain Boltzmann weights
for each update of the adiabatical 
fields through simple matrix  operations only.

\begin{figure}[htb]
\hfil\epsfxsize=8cm\epsfbox{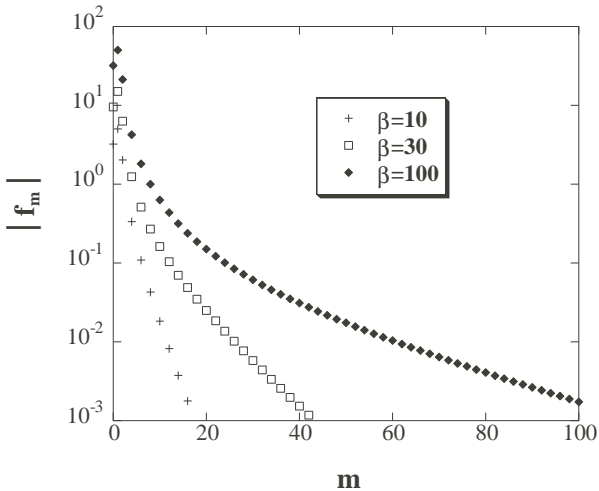}\hfil
\caption{Absolute values of the coefficient $|f_m|$ 
for the Chebyshev polynomial expansion
of $S_{\rm eff}$, for  $\beta=10$, $30$ and $100$ at $\mu=0$.}
\label{FigCoeff}
\end{figure}

Among various choices of polynomials
for the series expansion in Eq.~(\ref{defPolyExpand}),
the  Chebyshev polynomials\cite{Wang94,Silver94}
give us an advantage that the  expansion coefficients $f_m$
decay quickly  for  $m \gg 1$ in an exponential way.\cite{Motome99b}
The Chebyshev polynomials 
$\{T_m(x)\}$ for $m=0,1,\ldots$ are recursively defined by
\begin{eqnarray}
  T_0(x) &=& 1,\qquad 
  T_1(x) = x,  \nonumber\\
  T_m(x) &=& 2 x \,T_{m-1}(x) - T_{m-2}(x).
  \label{defChebyshev}
\end{eqnarray}
Within the region $-1 \le x \le 1$,
the Chebyshev polynomials show an orthonormal property 
 in a form
\begin{equation}
\int_{-1}^{1}\frac{{\rm d}x}
{\pi\sqrt{1-x^{2}}}
 T_m(x) T_{m'}(x) = \alpha_m \delta_{m{m'}},
\end{equation}
where 
\begin{equation}
\alpha_m = \left \{ 
  \begin{array}{ll}
    1, & m=0, \\
    1/2, \quad & m \ne 0.
  \end{array}
   \right. 
\end{equation}
Using this relation, the coefficients in Eq.~(\ref{defPolyExpand})
are obtained as
\begin{equation}
  f_{m} = \frac1{\alpha_m}
     \int_{-1}^{1}\frac{{\rm d}x}
     {\pi\sqrt{1-x^{2}}} F(x) T_{m}(x).
                               \label{defMoment}
\end{equation}
In Fig.~\ref{FigCoeff} we show some examples for 
absolute values of
the coefficients $|f_m|$ where we see  exponential decays.

From Eq.~(\ref{defmu}) we have 
\begin{eqnarray}
  \mu_m &=& 
   \int_{-1}^{1} {\rm d} \varepsilon D_{\mib\phi}(\varepsilon) T_m(\varepsilon)
   \nonumber\\
 &=& \int_0^\pi {\rm d}\theta \sin\theta D_{\mib\phi}(\theta)
  \cos m\theta,
  \label{defMuFourier}
\end{eqnarray}
where 
\begin{equation}
  D_{\mib\phi}(\varepsilon) 
   = \sum_\nu \delta(\varepsilon- \varepsilon_\nu(\mib\phi))
\end{equation}
is the density of states (DOS) for the
eigenvalues of $H(\mib\phi)$.
Here we used an alternative definition of the Chebyshev 
polynomials,
\begin{equation}
 T_m(\cos \theta)= \cos m \theta.
\end{equation}
We note here that, since Chebyshev polynomials $T_m(x)$ as
an orthonormal set are defined
in the region $-1 \le x \le 1$, the
Hamiltonian matrices have to be scaled properly so that
the eigenvalues satisfy $-1 \le \varepsilon_\nu(\mib\phi) \le 1$ 
for $\nu=1,\ldots, N_{\rm dim}$.
From Eq.~(\ref{defMuFourier}) 
we see that $\mu_m$ is a Fourier transform of 
$\sin\theta D_{\mib\phi}(\cos\theta)$.
This means that $|\mu_m|$ either  converge to zero  at $m \gg 1$
if the DOS is non-singular, or 
converge to a constant
in the most extreme cases where there exist macroscopic degeneracies 
in the DOS.
Therefore, the series $f_m \mu_m$ in Eq.~(\ref{CalcSeff}) decays
exponentially.

Replacement of
the infinite sum over $m$ in Eq.~(\ref{CalcSeff}) by
a finite sum up to $m=M$ gives us accurate results
if an appropriate value of $M$ is chosen.
Although such a value depends on an
asymptotic form of $\mu_m$ which is
not predictable {\em a priori}, we can always estimate
truncation errors of PEM  due to finite $M$
through comparisons between the results by the PEM and the DM.
Note that,  even when
it is difficult to perform a product run by the DM with 
large enough MC steps to obtain statistically accurate results,
it is usually possible to execute
 a few  trial MC steps in order to estimate truncation errors.
In MC runs in practice, 
truncation errors can be neglected if they are
sufficiently smaller than the statistical errors.

%%%%%:::::

\subsection{Algorithm}

The algorithm for the PEM 
using the Chebyshev polynomials is as follows.
We introduce a set of orthonormal vectors $\{\vec e(k) \}$ 
($k=1,\ldots,N_{\rm dim}$), and calculate
\begin{equation}
 \vec v(k,m) \equiv T_m(H) \,\vec e(k),
  \label{defVKM}
\end{equation} 
for $m\ge0$.
Then, the trace $\mu_m$
 is given by 
\begin{equation}
  \label{defTrace}
  \mu_m  =\sum_{k=1}^{N_{\rm dim}} \mu_m(k),
\end{equation}
where
\begin{eqnarray}
  \label{defMuK}
  \mu_{m}(k) &=&
  \left( \vec e(k) , \, \vec v(k,m)\right).
\end{eqnarray}
Here $(\ ,\ )$ represents an inner product.

Using a recursion relation for the Chebyshev polynomials
in Eq.~(\ref{defChebyshev}),
we obtain  $\vec  v(k,m) $ as 
\begin{eqnarray}
\vec  v(k,0)&=& \vec e(k),\qquad
\vec v(k,1)= H \vec v(k,0),
  \label{defProd01}
\end{eqnarray}
and 
\begin{equation}
  \label{defProdRecursive}
  v_i(k,m) 
   = 2\sum_{j} H_{ij} \, v_j(k,m-1) - v_i(k,m-2),
\end{equation}
for $i=1,\ldots,N_{\rm dim}$ at
$m\ge 2$, where
 $v_i(k,m)$ is the $i$-th element of the
vector $\vec v(k,m)$.

The summation $\sum_j$ in Eq.~(\ref{defProdRecursive}) can 
be restricted to $j$
with non-zero matrix elements $H_{ij} \ne 0$.
Since we assume sparse matrices in Eq.~(\ref{defProdRecursive}),
computational complexity of these matrix operations are
$O(N)$.
Therefore, for a fixed configuration of $\mib\phi$,
 the computational complexity to obtain $\mu_m(k)$ ($0\le m \le M$)
 is $O(MN)$, while that for the trace operation
is $O(N)$.
Then, the complexity of the
 Boltzmann weight calculation scales as $O(MN^2)$. This means 
that, for one MC step with local updates where 
$O(N)$ field variables are manipulated sequentially,
 the computational complexity is $O(MN^3)$.
Compared to the DM which scales as $O(N^4)$,
the PEM is advantageous if $M \ll N$.

It has also been shown that the algorithm 
is suited for parallel computations
since trace operations are mutually independent,\cite{Motome99b,Furukawa01a}
which further reduce elapsed time for calculations.
As a result, it become possible to investigate
models with larger system sizes within a reasonable scale of cpu time.
Using the PEM, 
critical phenomena at finite temperatures for a fermionic model 
 in two dimensions are studied for the first time
by finite-size scaling analysis
as well as by non-equilibrium
relaxation technique.\cite{Motome01,Motome01b}
However, 
the algorithm still turns out to be insufficient
to study critical phenomena in three dimensions,
since the reduction of the computational complexities
is not large enough.

%%%%%%%%%%%%%%%%%%%%%%%%%%%%%%%%%%%%%%%%%%%%%%%%%%%%%%%%%%%%%%%%%%%%%
\section{Truncated Polynomial Expansion Method}

Now we demonstrate that
the calculations 
for the Boltzmann weights
can further be  improved by introducing truncated matrix operations.
As an example to explain the method,
let us first consider a simple model 
which has the Hamiltonian matrix in the form
\begin{equation}
  H_{ij}(\mib\phi) = \left \{
     \begin{array}{ll}
       g \phi_i & i=j, \\
       -t             & \mbox{$i$ and $j$ are nearest neighbors,}\\
       0              &  \mbox{otherwise.}
     \end{array}
  \right. 
  \label{defHmatrixExample}
\end{equation}
Here $t$ is the nearest neighbor hopping energy for  spinless  electrons
while $g$ is the electron-field coupling constant.
In this system, local adiabatical field ${\mib\phi}= \{\phi_i \}$
is defined on each lattice which is coupled to electrons
as an on-site potential.
The Hamiltonian matrix $H({\mib\phi})$ 
is a sparse matrix with $N_{\rm dim}=N$.

%We hereafter assume that the PEM is successful in the sense 
%truncation errors due to the polynomial expansion up to
%the finite number of Chebyshev moments $M=O(N^0)$ are negligible.

\subsection{Truncation of matrix products}
In order to obtain $\mu_m(k)$ for $m=0,\ldots,M$ 
from Eqs.~(\ref{defMuK})-(\ref{defProdRecursive}),
matrix-vector multiplications
throughout the Hilbert space are necessary,
which give $O(MN)$ computational complexity.
Here we introduce a truncation of the matrix-vector multiplications
in order to reduce the computational complexity.

Let us choose  $ e_i(k)  =\delta_{ik}$ for the orthonormal set
in Eq.~(\ref{defVKM}).
Since $v_i(k,0)$ is non-zero only at $i=k$,
we have $v_i(k,1) \ne 0$ only at $i=k$ 
as well as at nearest neighbors of $k$.
Namely, 
due to the sparse nature of the  Hamiltonian matrix (\ref{defHmatrixExample}),
it is not necessary to calculate all the other vector elements.
Similarly, if one keeps track of a set of indices with
$v_i(k,m-1) \ne 0$, the calculation of
vector elements $v_i(k,m)$ can be restricted to
limited numbers of indices, so that the computational complexity is
much reduced.

\begin{figure}[htb]
\hfil\epsfxsize=8cm\epsfbox{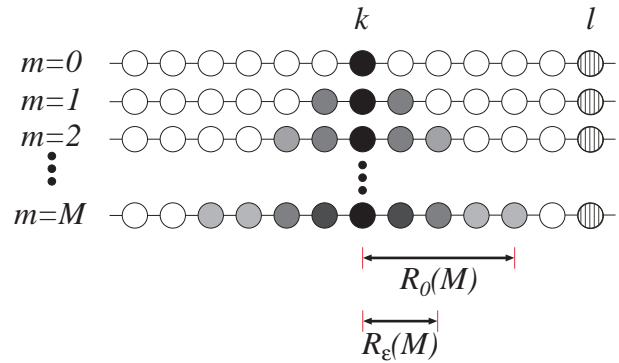}\hfil

\caption
{A  sketch for the propagation of vector elements.
The vertical axis gives the matrix-vector multiplication steps.
Circles aligned in the horizontal direction represent lattice sites.
Filled circles show sites with non-zero vector elements,
and darkness of them schematically
 illustrates  amplitudes of the vector elements.
 The initial vector  gives a localized state at site $k$,
while the hatched circle at $l$ symbolizes the site
where the Hamiltonian matrix element is updated.
See also the discussion in \S\ref{Subsec:trace}
}
\label{FigPropagate}
\end{figure}

The matrix-vector product in Eq.~(\ref{defProdRecursive}) can be
viewed as a transfer-matrix multiplication to a state vector,
which expresses a diffusive propagation of a wavefunction. 
We start from an initial  vector $\vec v(k,0)= \vec e(k)$,
which expresses an electron state localized at site $k$.
Each time the Hamiltonian matrix $H$ 
is multiplied to obtain $\vec v(k,m)$,
 electrons hop to nearest neighbors.
As a consequence,
 the sites with non-zero vector elements $v_i(k,m)$ spread out
as $m$ increases.
In Fig.~\ref{FigPropagate} we give a schematic illustration
for the propagation steps.
The process  also resembles the diffusion of the probability distribution
function  in a random-walk system.

Let us define the distance between $i$-th site and $k$-th site,
denoted as $||i-k||$,
by the minimum number of hops for an electron to
transfer from $i$-th site to $k$-th site.
On hypercubic lattices, this gives the ``Manhattan distance''.
We define the range of propagation $R_0(m)$ by
the longest distance that an electron can hop by $m$ steps
of the matrix-vector multiplication in Eq.~(\ref{defProdRecursive}).
Since sites which are outside of the range of propagation have
zero vector elements, {\em i.e.},
\begin{equation}
v_i(k,m) = 0  \mbox{\quad if $||i-k|| > R_0(m)$},
\end{equation}
we have no contribution to the calculations of $\mu_m(k)$ 
as well as $\vec v(k,m+1)$ from these sites.
This means that we may perform our matrix-vector calculation
only within the neighbors of $k$ which 
satisfies $||i-k|| \le R_0(m)$ to obtain exact results.
In the present model (\ref{defHmatrixExample}) 
we have  $R_0(m)=m$, and the number of sites
which contributes to the overall calculation of $\mu_m(k)$ ($m\le M$) 
is proportional to $M^d$ instead of $N$ when every sites on the lattice
is taken into account.
Here $d$ is the spacial dimension of the lattice.
By introducing this restriction, 
the computational complexity to obtain $\mu_m(k)$ ($m\le M$) 
is reduced from
$O(MN)$ to $O(M^{d+1})$ without any cost for the computational accuracies.

We can further reduce the range where calculations are restricted,
 by introducing a threshold $\epsilon$ for the vector elements.
When the absolute values of the vector elements 
$|v_i(k,m)|$ are small enough, we can neglect
such terms in the calculation of $\mu_m(k)$ and $\vec v(k,m+1)$.
Let us define the range of propagation $R_\epsilon(m)$ by
the longest distance $||i-k||$ such that the absolute value
of the vector element on the $i$-th site exceeds the threshold,
$|v_i(k,m)| \ge \epsilon$.
In Fig.~\ref{FigPropagate} we give a schematic illustration.
Then, we have
\begin{equation}
 |v_i(k,m)| < \epsilon  \mbox{\quad if $||i-k|| > R_\epsilon(m)$},
\end{equation}
and contributions to $\mu_m(k)$ from sites outside of the range
are negligible if we take  values of $\epsilon$
appropriately.

Since the diffusion length 
is proportional to the square-root of the time-steps in general, 
$R_\epsilon(m) \propto \sqrt{m}$ for $\epsilon >0$.
Then, by introducing a threshold $\epsilon$ and
restricting the calculation
within $||i - k || \le R_\epsilon(m)$,
the number of sites which contributes to $\mu_m(k)$ 
scales as $O(M^{d/2})$. 
Therefore
the computational complexity to obtain 
$\mu_k(m)$ ($m\le M$) is further reduced to $O(M^{d/2+1})$
with the accuracies of $O(\epsilon)$.
(To be more specific, the number of sites neglected by this
treatment is  $O(M^d)$
for the calculation of $\mu_m$ ($m=0,\ldots,M$), 
so the total error for the calculation
of the Boltzmann weight is $O(M^{d+1}\epsilon)$.)

Let us now extend the procedure to general cases
where an index for fermion degrees of freedom
represents a combination of site, orbital and spin. 
Since the basis set to define the Hamiltonian matrices (\ref{defHgeneral})
can be taken arbitrary,
we may also consider a system where it is not well-defined to
consider a geometrical distance between indices.
Nevertheless, as long as the Hamiltonian matrices are sparse,
we can generalize the algorithm as follows.

We define 
a subspace ${\cal N}_0(k,m)$ as a set of neighboring indices of
the initial index $k$ which are 
within the range of propagation by $m$ steps,
\begin{equation}
  {\cal N}_0(k,m) = \bigcup_{m'=0}^m \{ i \ |\  v_i(k,m')\ne 0 \}.
\end{equation}
Note that, in the previous example,
 $ {\cal N}_0(k,m)$ is a set of indices within
the range of $R_0(m)$ from $k$.
Then, we may restrict the matrix product operations 
within the subspace ${\cal N}_0(k,m)$,
since 
\begin{equation}
  v_i(k,m)= 0   \mbox{\quad if $ i \not\in {\cal N}_0(k,m)$},
\end{equation}
so that there is no contribution
to $\mu_m(k)$ from outside of ${\cal N}_0(k,m)$.

Similarly,
we define a subspace ${\cal N}_\epsilon(k,m)$ as a set of 
neighboring indices of $k$
where absolute values of the vector elements exceed the threshold 
$\epsilon$,
\begin{equation}
  {\cal N}_\epsilon(k,m) 
  = \bigcup_{m'=0}^m \{ i \ |\  |v_i(k,m')|> \epsilon \}.
\end{equation}
In the previous example, $ {\cal N}_\epsilon(k,m)$ 
roughly corresponds to a set of sites within
the range of $R_\epsilon(m)$ from $k$.
By making truncations of the matrix product operations 
within the restricted subspace ${\cal N}_\epsilon(k,m)$,
we obtain approximate results for
$\mu_m(k)$ within  errors of $O(\epsilon)$,
since 
\begin{equation}
 |v_i(k,m)| < \epsilon  
    \mbox{\quad if $i \not\in {\cal N}_\epsilon(k,m)$},
\end{equation}
so that contributions from outside of ${\cal N}_\epsilon(k,m)$ are
negligible.

%%%%%%%%%%%%%%%%%%%%%

\subsection{Truncation of trace operations}
\label{Subsec:trace}
In order to obtain $S_{\rm eff}$ directly
from Eqs.~(\ref{CalcSeff}) and (\ref{defmu}),
a trace operation throughout the Hilbert space is necessary,
which gives $O(N)$ computational complexity.
Here we introduce a truncation of the trace operation
in order to reduce the computational complexity.

The probability of the MC update
from an old field configuration ${\mib\phi^{\rm old}}$
to a new configuration ${\mib\phi^{\rm new}}$, 
which is given by
the ratio of the Boltzmann weights in Eq.~(\ref{defBoltzRatio}),
can be calculated from
\begin{equation}
  \frac{P({\mib\phi^{\rm new}})}{P({\mib\phi^{\rm old}})}
 = \exp( -\Delta S_{\rm eff}),
\end{equation}
where $\Delta S_{\rm eff}$ is the difference of the
effective action.
Using PEM up to the $M$-th order, we have
\begin{eqnarray}
  \label{defDeltaSeff} 
\Delta S_{\rm eff}&=& S_{\rm eff}(\mib\phi^{\rm new})-
S_{\rm eff}(\mib\phi^{\rm old})
  \nonumber\\
&\simeq& \sum_{m=0}^M f_m \sum_{k}^{N_{\rm dim}}  \Delta\mu_m(k),
  \label{DeltaSeffTrace}
\end{eqnarray}
where
\begin{eqnarray}
 \Delta\mu_{m}(k) &=& 
(\vec e(k), \vec v^{\rm new}(k,m)) \nonumber\\ 
&&\quad 
 - 
   (\vec e(k), \vec v^{\rm old}(k,m)).
\end{eqnarray}
Here, $\vec v^\alpha (k,m)$ for $\alpha= ({\rm old}, {\rm new})$
is defined by
\begin{equation}
  \label{defVecNewOld}
  \vec v^\alpha(k,m) = T_m(H(\mib\phi^\alpha)) \vec e(k),
\end{equation}
and we choose $e_i(k) = \delta_{ik}$.
Summation over $k$ in Eq.~(\ref{defDeltaSeff}) is the 
trace operation in Eq.~(\ref{defmu}).

Now we consider a local update of the adiabatical fields
in the present exemplified model (\ref{defHmatrixExample}).
Let us choose a site $l$
and try a local update on the site
$\phi_{l}^{\rm old} \to \phi_{l}^{\rm new}$, while
 we have $\phi_{i}^{\rm old} = \phi_{i}^{\rm new}$
for $i\ne l$.
In this case, the change of the Hamiltonian matrix
$H(\mib\phi^{\rm old}) \to H(\mib\phi^{\rm new})$ exists
only at the $l,l$-th matrix element, while
we have $H_{ij}(\mib\phi^{\rm old}) = H_{ij}(\mib\phi^{\rm new})$
elsewhere.

Let us take a site $k$ which is distant enough from the
updated site $l$ so that $||k-l|| > R_0(M)$ is satisfied,
and consider the diffusion of the vectors $ \vec v^{\rm old}(k,m)$
and $ \vec v^{\rm new}(k,m)$.
In this case,  we have $\vec v^{\rm old}(k,m)=\vec v^{\rm new}(k,m)$
and hence $\Delta \mu_m(k)=0$.
The reason is as follows.
For $m \le M$, the region where the state vectors
propagate does not reach the site $l$, since
 $ R_0(m) < ||k-l||$ is satisfied.
In Fig.~\ref{FigPropagate} we give a schematic illustration.
The matrix elements that are operated to the vectors
during the diffusion processes
are identical between old and new configurations.
This makes
\begin{equation}
  v_i^{\rm old}(k,m) =   v_i^{\rm new}(k,m)  
   \label{EqVdiffIn}
\end{equation}
for $i$ which satisfies  $||i-k|| \le R_0(m)$.
At the same time, by the definition of $R_0(m)$ we have
$  v_i^{\rm old}(k,m) = 0$ and $  v_i^{\rm new}(k,m)=0$
for $i$ such that   $||i-k|| > R_0(m)$.
Then,
we have  $ v_i^{\rm old}(k,m) =   v_i^{\rm new}(k,m)$ in the
entire space.

In other words, we have $\Delta \mu_m(k) \ne 0$ for $m\le M$
only if $k$ is close enough to $l$ so that the
propagation from the site $k$ reaches the site $l$ within $M$ steps.
Therefore, it is sufficient to take the summation
over $k$ in Eq.(\ref{DeltaSeffTrace}) only 
within the vicinity of $l$  which satisfies $||k-l|| \le R_0(M)$.
Namely, the trace operation may be restricted to a subspace
which has $O(M^d)$ sites so that the computational complexity
for the trace operations
is reduced from $O(N)$ to $O(M^d)$.

Furthermore, we introduce a truncation of trace operations
which gives approximate results with a reduced computational complexity.
Here we consider a general case with sparse
Hamiltonian matrices. For a moment
we restrict ourselves
to an update of the adiabatical field where the matrix elements
of $H(\mib\phi^{\rm old})$ and $H(\mib\phi^{\rm new})$
are identical except for the $l,l$-th element.

Let us consider an initial vector at $k$ and the
propagation of the vectors for $H(\mib\phi^{\rm old})$ and 
$H(\mib\phi^{\rm new})$,
which gives the set of neighboring indices 
${\cal N}_\epsilon^{\rm old}(k,m)$
and  ${\cal N}_\epsilon^{\rm new}(k,m)$, respectively.
If $l \notin {\cal N}^{\rm new}_\epsilon(k,M)$ and 
$l \notin {\cal N}^{\rm old}_\epsilon(k,M)$ are satisfied,
both $\vec v^{\rm old}(k,m)$ and $\vec v^{\rm new}(k,m)$ 
are approximately confined within the subspace where matrix elements of the
Hamiltonian are identical, and do not reach the index $l$. 
Then we have 
\begin{equation}
 \vec v^{\rm old}(k,m) \simeq \vec v^{\rm new}(k,m),
\end{equation}
within the error threshold $\epsilon$, and therefore
$ \Delta\mu_m(k)= O(\epsilon)$ is satisfied.
In other words, only indices in the vicinity of $l$
where effective propagations to $l$ occur within $M$ steps
should be considered for the calculations of $\Delta\mu_m(k)$.

This means that the trace operation in Eq.~(\ref{defDeltaSeff}) can be
restricted to the vicinity  of $l$,
defined by a set of indices $k$ where 
$ l \in {\cal N}^{\rm new}_\epsilon(k,M)$ or
$l \in {\cal N}^{\rm old}_\epsilon(k,M)$ are satisfied.
Using Eq.~(\ref{defVecNewOld})
and the Hermicity of the Hamiltonian matrix polynomials
\begin{equation}
 \left. \vphantom{T_M^M} T_m(H) \right|_{lk} 
  = \left(\left. \vphantom{T_M^M}T_m(H) \right|_{kl}\right)^*,
\end{equation} 
we have $v_l^\alpha(k,m) = v_k^\alpha(l,m)^*$ 
for $\alpha = ({\rm old}, {\rm new})$.
Namely, if and only if $k \in N_\epsilon^\alpha(l,M)$, 
we have $l \in N_\epsilon^\alpha(k,M)$.
Therefore,
the trace operation can be restricted to a subspace defined by 
\begin{equation}
  {\cal V}_\epsilon(l,M) \equiv {\cal N}^{\rm old}_\epsilon(l,M) 
  \cup {\cal N}^{\rm new}_\epsilon(l,M).
  \label{defVisinity}
\end{equation}
Due to the diffusive nature of the propagation,
the number of indices in ${\cal N}^{\rm old}_\epsilon(l,M)$ 
and  ${\cal N}^{\rm new}_\epsilon(l,M)$ is $O(M^{d/2})$
on usual lattice systems.
The truncated trace operation within ${\cal V}_\epsilon$ reduces the
computational complexity from $O(N)$ to  $O(M^{d/2})$
with errors of $O(\epsilon)$.

We can extend our algorithm to cases where a local update modulates
matrix elements for multiple indices.
An example is the case where fields are coupled to
off-diagonal matrix elements, {\em e.g.}, hopping amplitudes.
Let us consider a case where
a local update  modulates the  $l,l'$-th matrix element.
Differences between $\vec v^{\rm old}$ and $\vec v^{\rm new}$
exist if the propagations of the vectors reach
either of  the indices $l$ or $l'$.
Then, we need to consider a sum of vicinities
centered at $l$ and $l'$,
\begin{equation}
  {\cal V}_\epsilon^{\rm tot}= 
   {\cal V}_\epsilon(l,M) \cup {\cal V}_\epsilon(l',M),
\end{equation}
and make trace operations within ${\cal V}_\epsilon^{\rm tot}$.
In a general case where a number of matrix elements
are modulated, we need to consider all the indices
associated with modulated matrix elements.
We define  
${\cal C}$ as a set of indices where matrix elements are
modulated by the update,
\begin{equation}
  {\cal C} = \{l \ | \ {}^\exists l',\  
  H_{ll'}(\mib\phi^{\rm old}) \ne H_{ll'}(\mib\phi^{\rm new}) \}.
  \label{defChangeMatrixElements}
\end{equation}
Then,
the total vicinity $ {\cal V}_\epsilon^{\rm tot}$ is given by
\begin{equation}
    {\cal V}_\epsilon^{\rm tot}= 
      \bigcup_{l \in {\cal C}}  {\cal V}_\epsilon(l,M),
  \label{defVisinityTotal}
\end{equation}
and the  trace operations are performed within $ {\cal V}_\epsilon^{\rm tot}$,
$\em i.e.$,
\begin{equation}
\Delta S_{\rm eff}
   \simeq \sum_{m=0}^M f_m 
          \sum_{k \in {\cal V}_\epsilon^{\rm tot}}  \Delta\mu_m(k).
  \label{defDeltaSeffTruncate}
\end{equation}
As long as MC updates are local, the number of indices in
${\cal C}$ is $O(N^0)$, so that the computational complexity 
for the trace operation will be $O(M^{d/2})$.

\subsection{Comparison with previous methods}

Thus we see that the PEM using truncated matrix operations
reduces the total computational complexity for one local update
from $O(M N^2)$ to $O( M^{d+1})$,
by combining 
threshold truncations for both matrix products and trace operations.
Hereafter we refer to the improved method in this section
with truncations of matrix operations as the truncated PEM,
whereas the original method described in \S\ref{Section:PEM} is 
called as the full PEM.
In Table~\ref{TableOrderN} we summarize the computational
complexities for various algorithms for comparison.

\begin{table}[ht]
\caption{
 Computational complexities to perform calculations
of $\mu_m(k)$, trace operations, calculations of the Boltzmann
weight ratio through $\Delta S_{\rm eff}$, and 
a  MC step with local updates in total.
Here, f-PEM and t-PEM stand for the full PEM and the truncated PEM,
respectively. Threshold for the truncated PEM is described by
$\epsilon$.}

\begin{tabular}{@{\hspace{\tabcolsep}\extracolsep{\fill}}ccccc}
\hline
Algorithm & $\mu_m(k)$  & Trace & 
$\Delta S_{\rm eff}$ & Total \\
\hline
DM & -- & -- & $O(N^3)$ & $O(N^4)$ \\
f-PEM & $O(MN)$ & $O(N)$ & $O(MN^2)$  & $O(MN^3)$ \\
t-PEM &&&&\\
$\epsilon=0$
& $O(M^{d+1})$ & $O(M^d)$ & $O(M^{2d+1})$  & $O(M^{2d+1}N)$ \\
$\epsilon\ne0$
& $O(M^{\frac{d}2+1})$ & $O(M^{\frac d2})$ & $O(M^{d+1})$  & $O(M^{d+1}N)$ \\
\hline
\end{tabular}
\label{TableOrderN}
\end{table}

Let us emphasize here
that the restriction of matrix operations
within ${\cal N}_0(k,M)$ gives us  identical results for
$\mu_m(k)$ to those obtained by the full PEM,
 with a reduced computational complexity.
We also have ${\cal N}_{\epsilon\to0}(k,M)={\cal N}_0(k,M)$.
This implies that the introduction of the threshold $\epsilon$
is a controlled approximation in the sense that 
$\mu_m(k)$ are obtained with an arbitrary accuracy by an appropriate
choice of $\epsilon$, with further reduced computational complexities.

\section{Algorithm and Implementation}

\subsection{Algorithm}

Now we clarify  actual algorithms to perform the truncated PEM.
We first show an algorithm for the truncated matrix product
 to obtain $v_i(k,m)$ for $m\le M$:
\begin{itemize}
\item[i)]  Determine
  the truncation threshold for matrix products $\epsilon_{\rm p}$.
 Set the initial unit vector $v_i(k,0) = \delta_{ik}$, and the
 initial restricted subspace ${\cal N}_{\epsilon_{\rm p}} = \{k\}$.
\item[ii)]  From given $\{v_i(k,m-1)\}$ and ${\cal N}_{\epsilon_{\rm p}}$,
 perform matrix-vector product   to  calculate $\{v_i(k,m)\}$
 within restricted subspace.
 Namely,  indices $j$
  in Eq.~(\ref{defProdRecursive}) are restricted to 
  $j \in {\cal N}_{\epsilon_{\rm p}}$, whereas indices $i$ also
  include those generated by propagations due to 
  non-zero off-diagonal matrix elements
  of $H_{ij}$ in Eq.~(\ref{defProdRecursive}).
\item[iii)] The treatment for a newly generated index $i$ is as follows:
   If $|v_i(k,m)| \ge \epsilon_{\rm p}$, 
  register $i$ as a new member of ${\cal N}_{\epsilon_{\rm p}}$.
  Otherwise, discard the calculation for $v_i(k,m)$ and treat it as zero. 
\item[iv)] Repeat steps ii) and iii) $M$ times.
\item[v)] As a byproduct, we obtain ${\cal N}_{\epsilon_{\rm p}}(k,M)$.
\end{itemize}
This procedure automatically gives calculations without truncation errors
by setting $\epsilon_{\rm p}=0$.

An algorithm to perform truncated trace operation for
an update of the field is as follows:
\begin{itemize}
\item[i)]  Determine the truncation  threshold
for trace operations $\epsilon_{\rm tr}$.
\item[ii)] Define ${\cal C}$ as in Eq.~(\ref{defChangeMatrixElements}).
Perform truncated matrix product operations
 to obtain $\vec v^\alpha(l,M)$  for  $l \in {\cal C}$ and
 $\alpha=({\rm new}, {\rm old})$, so that 
 the neighbors of the local-updated indices
 ${\cal N}_{\epsilon_{\rm tr}}^\alpha(l,M)$ are determined.
 Then,    ${\cal V}_{\epsilon_{\rm tr}}^{\rm tot}$ is obtained
 from Eqs.~(\ref{defVisinity}) and (\ref{defVisinityTotal}).
\item[iii)] For $k \in {\cal V}_{\epsilon_{\rm tr}}^{\rm tot}$,
calculate $\vec v^{\rm old}(k,M)$ and $\vec v^{\rm new}(k,M)$,
and obtain $\Delta\mu(k)$.
Finally,  $\Delta S_{\rm eff}$
is evaluated from Eq.~(\ref{defDeltaSeffTruncate}).
\end{itemize}
This procedure also gives error-free results
if $\epsilon_{\rm tr}=0$.

Let us note that,
the thresholds $\epsilon_{\rm p}$ and $\epsilon_{\rm tr}$
may be chosen independently, and
the choice of these thresholds has to be justified by
making proper estimates for the truncation errors.

Although there exists an arbitrary choice
 for the basis set to express the system
in a quadratic form by Hamiltonian matrices (\ref{defHgeneral}),
one should select that which
reduces the propagation of the state vector as much as possible
in order to reduce the computational complexities.
A possible example is the case where there exists a symmetry
in a system so that the Hamiltonian matrices may be
block-diagonalized. The symmetry can  be weakly broken as long as
matrix elements between blocks are small.  Then, using the basis set
which block-diagonalize the Hamiltonian matrices,
the vector propagations will be confined within the blocks.

Similarly, 
one may also make an extention to the algorithm
so that the procedure for matrix products
starts from an initial unit vector $v_i(k,0) \ne \delta_{ik}$,
provided the number of the non-zero vector elements in
$\vec v(k,0)$ is $O(N^0)$.
An appropriate choice of the initial vector,
typically an expression of local symmetries of the system,
may cancel the propagation to some extent through interferences.

%%%%%%%%%:::::::::::

\subsection{Benchmark}
In order to demonstrate that the algorithm is successfully
implemented, we show benchmark results.
As a model, we choose the double-exchange (DE) model
in the strong-coupling limit in three dimensions.
The Hamiltonian is given by
\begin{equation}
  \hat {\cal H}(\{\vec S_i\}) = - \sum_{<ij>}
  t(\vec S_i, \vec S_j) (c_{i}^\dagger  c_{j} +h.c.),
   \label{defHDE}
\end{equation}
where  hoppings of spinless electrons to nearest neighbors
are coupled to classical spin fields $\{\vec S_i\}$
in a form
\begin{equation}
\frac{t(\vec S_i,\vec S_j)}{t_0}=
  \cos\frac{\theta_i}2 \cos\frac{\theta_j}2
   + \sin\frac{\theta_i}2 \sin\frac{\theta_j}2 
   e^{ -{\rm i} (\phi_i-\phi_j)}.
\end{equation}
Here $\theta$ and $\phi$ are defined by the direction of
the localized spin $\vec S$ as
\begin{equation}
  S_i^x = S \sin \theta_i \cos \phi_i, \ 
  S_i^y = S \sin \theta_i \sin \phi_i, \ 
  S_i^z = S \cos \theta_i,
\end{equation}
with the normalization  $S=1$, while
$t_0$ is the transfer integral between
nearest neighbors in the absence of the
DE interaction.
A local update for $\vec S_i$ modulates
all the hopping energies from $i$-th site to its nearest neighbor sites.
Then, $\cal C$ for  site $i$  contains nearest neighbors
of $i$ as well as $i$ itself.

For parameters of the MC benchmark run, we choose
$T/W\sim 0.02$ and $\mu/W\sim 0$ where $T$ and $\mu$ are temperature and
the chemical potential, respectively, while $W=6t_0$ is the half-bandwidth
of the model in the absence of the interactions.
We typically take MC steps as $N_{\rm step}\sim 4000$.
These parameters are 
typical ones for an investigation of the ferromagnetic transition
in the model.\cite{Motome00we}
Within this parameter range,
$M=16$ has been shown to be
enough for the accuracy of the calculation.

\begin{figure}[htb]
\hfil\epsfxsize=8cm\epsfbox{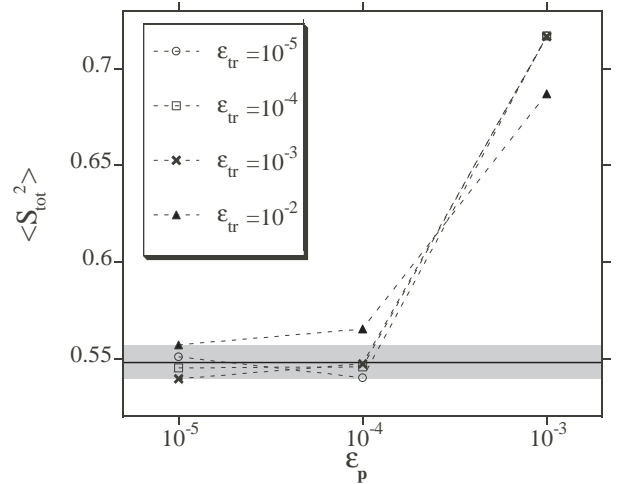}\hfil
\caption{Deviations of the estimated values due to
truncations, for the square of the
  ferromagnetic moment
 $S_{\rm tot}{}^2$ on a $4\times4\times6$ lattice system
at $T/W=0.015$, $\mu/W=0$, $M=16$ and $N_{\rm step}=4000$.
Symbols show estimates by the truncated PEM
for various values of the truncation thresholds
$\epsilon_{\rm p}$ and $\epsilon_{\rm tr}$. The solid line in the figure
shows the estimated value of $S_{\rm tot}{}^2$ obtained by the DM,
 while the gray area around the line
gives the stochastic error bar for the estimate. Error bars of each symbols
are roughly equal to the error bar of the data by the DM.
Therefore, if symbols are in the gray area, it is conceivable that
overall truncation errors are roughly 
equal to or smaller than stochastic errors.}
\label{FigError}
\end{figure}

In Fig.~\ref{FigError}, we show truncation errors for
various combinations of  $\epsilon_{\rm tr}$ and
$\epsilon_{\rm p}$. 
We see that the truncation errors quickly decreases as
the thresholds are lowered, and becomes sufficiently small
compared to the statistical errors.
From this result, we choose $\epsilon_{\rm tr}= 10^{-3}$
and $\epsilon_{\rm p}=10^{-5}$ which are 
satisfactory to give accurate results with reduced cpu time
in the present case.

In usual cases, it is expected that we may take
 $\epsilon_{\rm tr}$  larger
than $\epsilon_{\rm p}$. 
The threshold $\epsilon_{\rm tr}$ 
determines the border of the region
  ${\cal V}_{\epsilon_{\rm tr}}^{\rm tot}$.
For the indices at the border,
contributions to $\Delta \mu_m(k)$ 
come from propagations from these indices to those in ${\cal C}$,
which take small fractions of the whole propagations.
On the other hand, 
the threshold $\epsilon_{\rm p}$ 
determines the border of all the propagations,
including those from the indices near ${\cal C}$
which give relatively large values of $\Delta \mu_m(k)$.
Therefore, $\Delta S_{\rm eff}$ is more sensitive to 
the threshold $\epsilon_{\rm p}$.

\begin{figure}[htb]
\hfil\epsfxsize=8cm\epsfbox{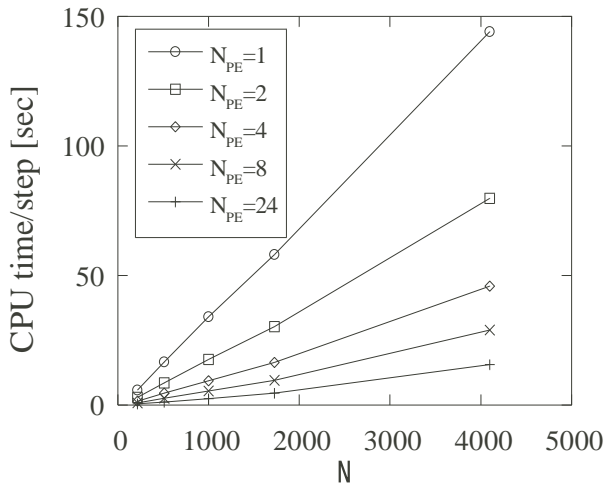}\hfil
\caption{Benchmark results for $6\times6\times6$, $8\times8\times8$,
$10\times10\times10$, $12\times12\times12$ and $16\times16\times16$ sites
of the double-exchange model in three dimensions,
at $T/W=0.02$, $\mu/W=0$ and $M=16$, while $\epsilon_{\rm tr}= 10^{-3}$
and $\epsilon_{\rm p}=10^{-5}$ are chosen for the thresholds.
Computations are
performed up to $N_{\rm PE}=24$ processor elements of an Athron MP1500+
cluster system which are connected by Myrinet 2000.}
\label{FigBench}
\end{figure}

Figure \ref{FigBench} shows the cpu time per MC step
as a function of the system size
on a single processor system as well as  on a parallel computational
system with the number of processor elements $N_{\rm PE} \le 24$.
From the figure we 
confirm that the cpu time is proportional to the system size.
We also see that it is possible to implement
this algorithm for efficient parallel computations.

Table \ref{TableBenchAthron} gives  benchmark results of the truncated
PEM for various system sizes, in comparison with
the diagonalization method and the full PEM.
Using the truncated PEM, it becomes possible to calculate
large size systems in a realistic time scale. 
For example, DE model in three dimensions
up to $20\times20\times20$ becomes available,
which is large enough to perform a finite-size scaling analysis
of the critical phenomena in this model.
The result is reported elsewhere.\cite{Motome03b}

\begin{table}[ht]
\caption{
An estimated cpu time of $10,000$ MC steps for various system sizes of
the  DE model
in three dimensions, at $T/W=0.02$ and $\mu/W=0$.
For the PEM we have $M=16$, while $\epsilon_{\rm tr}= 10^{-3}$
and $\epsilon_{\rm p}=10^{-5}$ are chosen for the 
truncation thresholds.
The cpu time is estimated
on a personal computer with a single processor of
the Athron MP1500+ processor. 
}
\begin{tabular}{@{\hspace{\tabcolsep}\extracolsep{\fill}}cccc}
\hline
System size & Diagonalization  & Full PEM & Truncated PEM\\
\hline
$8\times8\times8$ & $2.3$ years & $82$ days & $2.4$ days\\
 $12\times12\times12$ & $300$ years & $8.7$ years & $8$ days\\
 $16\times16\times16$ & $9500$ years & $120$ years & $21$ days\\
\hline
\end{tabular}
\label{TableBenchAthron}
\end{table}

\section{Summary and Discussions}

Under the condition that the full PEM
reduces the computational complexity for one 
MC step with local updates to $O(N^3)$ from that for the
DM which is $O(N^4)$, namely,
if  we may take $M=O(N^0)$ to obtain accurate results,
we have shown that the truncated PEM  further reduces it to $O(N)$.
It is possible to obtain the exact results within the PEM scheme 
if we take
$\epsilon_{\rm p}=\epsilon_{\rm tr}=0$.
The computational complexity can further be reduced by
introducing non-zero values for
$\epsilon_{\rm p}$ and $\epsilon_{\rm tr}$.
The truncation
errors can be made arbitrary small by taking small enough
values for $\epsilon_{\rm p}$ and $\epsilon_{\rm tr}$.

So far we have restricted ourselves to systems 
with sparse Hamiltonian matrices as well as short-range interactions
between adiabatical fields and electrons.
When interactions are long ranged while  the Hamiltonian 
matrices are still sparse, 
the PEM is applied with less efficiencies.
A local update of the adiabatical fields
modulates large numbers of Hamiltonian matrices, and therefore
the number of indices in
 ${\cal V}_{\epsilon}^{\rm tot}$ may be proportional to $N_{\rm dim}$
in the procedure of
the truncation of the trace operation.
The truncation
of the matrix-vector multiplication works similarly as before.
Then, the  computational complexity for a local update of the fields
scales as $O(M^{\frac d2 + 1}N)$ instead of $O(M^{d+1})$.
Similar results will be obtained if one performs
a global update of the adiabatical fields
where number of the fields to be updated is $O(N)$.

However,
in the case of systems with dense matrices,
the PEM is completely inefficient.
Namely, a multiplication of the Hamiltonian matrix make the vector
to propagate to the whole space,
so that procedures to restrict the subspace for calculations
 do not reduce computational complexities.
Moreover, the matrix-vector multiplications cost
$O(N_{\rm dim}^2)$ instead of $O(N_{\rm dim})$.
The full PEM as well as the truncated PEM gives $O(MN^4)$
 computational complexities,
and in this case, the DM should simply be used for calculations.

If we consider a case where calculations are performed
on small size lattices, 
 the propagation of the vectors
quickly spreads to the whole space. 
In other words, the Hamiltonian matrices effectively
become dense. In this case, the truncated PEM 
does not reduce the computational complexities in practice.
In general, for each system 
there exists a minimum number for system sizes where
the truncated PEM is avdantageous compared to the DM.
The minimum number depends on 
the range of the vector propagations with $M$-steps,
determined by model parameters as well as  lattice dimensions
and geometries. 

Near the critical points, improved MC sampling techniques such
as the histogram method or the multicanonical method
are used to overcome the limitations of the importance
sampling MC method.\cite{Landau00}
These techniques can also be applied to the 
truncated PEM, where the energy of a sample is given by
the ``effective electronic free energy'' defined by
\begin{equation}
   F_{\rm eff}(\mib\phi)  = S_{\rm eff}(\mib\phi)/\beta.
\end{equation}

Let us finally 
comment on the fact that $M$ may be kept to a constant which means that
the propagations of electrons which contribute to $S_{\rm eff}$
 are limited to a finite distance $R_0(M)$.
The situation is valid even at a critical point
where correlation length for the classical fields diverges,
since $S_{\rm eff}$ are determined
not by correlations but by the effective interaction
energies among classical fields which may be short ranged.
From the other point of view, 
 actions are in general non-singular at critical points
so that the polynomial expansion  converges stably.

\section*{Acknowledgment}

Calculations are performed on
a AOYAMA+ system  ({\tt http://www.phys.aoyama.ac.jp/$^\sim$aoyama+/}),
parallel Athron MP1500+ PC computers connected by Myrinet 2000.
The authors thank H. Nakata and M. Tsutsui for helpful supports in
developing parallel-computational systems.
This work is supported by a Grant-in-Aid from the
Ministry of Education, Culture, Sports, Science and Technology.

%%%%%%%%%%%%%%%%%%%%%%%%%%%%%%%%%%%%%%%%%%
%\bibliographystyle{jpsj}
%\bibliography{all,local}

\end{document}